\documentstyle[psfig,amssymb,amsfonts]{aa}

\def\3he{$^3$He}
\def\12sur13{$^{12}$C/$^{13}$C}
\def\Msun{M$_{\odot}$}
\newcommand{\nc}{\newcommand}
\nc{\ALi}{$A_{\rm Li}$\,}
\nc{\Teff}{$T_{\rm eff}$\,}
\nc{\logg}{log~$g$~}
\nc{\kms}{\,${\rm km.s}^{-1}$}
\nc{\Ciso}{${\rm ^{13}C}$}
\nc{\C}{${\rm ^{12}C}$}
\nc{\Cratio}{\C~/~\Ciso}

\begin{document}
\thesaurus{ 02.03.3;  
            08.01.1;  
	    08.05.3;  
	    08.09.3;  
            08.12.1;  
            08.18.1.  
          }

\title{ Lithium and rotation on the subgiant branch}
\subtitle {II. Theoretical analysis of observations}

\author{J. D. do Nascimento Jr, \inst{1,}\inst{4}
        C. Charbonnel, \inst{1}  
        A. L\`ebre,    \inst{2}
        P. de Laverny, \inst{3}
       J.R. De Medeiros  \inst{4}
                 }
\offprints{J. D. do Nascimento Jr (dias@obs-mip.fr)} 

\institute{Laboratoire d'Astrophysique de Toulouse, UMR 5572 CNRS, 16 Avenue
E. Belin, 31400 Toulouse, France 
               \and 
        GRAAL, UPRES--A 5024 CNRS, CC 072, Universit\'e Montpellier II,
         F--34095 Montpellier Cedex, France
                \and
         Observatoire de la C\^ote d'Azur, Fresnel, UMR6528, BP4229,
          06304 Nice CEDEX 4, France
                   \and  
Departamento de F\'{\i}sica, Universidade Federal do Rio Grande do Norte, 59072-970  Natal,
R.N., Brazil }

\date{Received April 1999, Accepted XX 1999}  

\maketitle

\markboth{J. D. do Nascimento Jr {\it et al.} :  Lithium and rotation on the
subgiant branch. II}{J. D. do Nascimento Jr {\it et al.} : Lithium and rotation on
the subgiant branch. II }

\begin{abstract}
{\sf 
Lithium abundances and rotation, determined for
120  subgiant stars in L\`ebre {\it et al.} (1999) are analyzed.
To this purpose, the evolutionary status 
of the sample as well as the individual masses have been determined
using the HIPPARCOS trigonometric parallax measurements  to locate very
precisely our sample stars in the HR diagram. We look at the distributions
of \ALi and {\it Vsini} with mass when  stars evolve from the main sequence
to the subgiant branch.

For most of the
stars in our sample we find  good agreement with the dilution predictions. 
However, the more massive cool stars 
with upper limits of Li abundances show a significant discrepancy  with the 
theoretical predictions, even if the Non-LTE effects are taken into account.
For the rotation behaviour, our analysis 
 confirms
that low mass stars leave
the main sequence with a low rotational rate, while more massive stars
are slowed down only when reaching the subgiant branch. We also checked
the connection between the observed rotation behaviour and the magnetic braking
due to the deepening of the convective envelope. Our results shed new
light on the lithium and rotation discontinuities in the evolved phase.

}
\keywords{convection -- stars: abundances -- stars: evolution --
          stars: interiors -- stars: late-type  -- stars: rotation
}

\end{abstract}

\section{Introduction}

This paper is the second in a series about the study of lithium and 
rotation in evolved stars based on both new high resolution spectroscopic
observations and precise rotational velocities obtained with the
CORAVEL spectrometer. 
In L\`ebre {\it et al.} (1999, hereafter Paper I) we derived lithium
abundances by spectral synthesis analysis and presented the observational data 
for a large and homogeneous sample of F, G, and K-type Population I 
subgiant stars. On the basis of this uniform data set, we could confirm 
the occurrence of the rotational discontinuity near the spectral type 
F8IV (Gray \& Nagar 1985, De Medeiros \& Mayor 1989, 1990),
and localize a lithium drop-off around the spectral type G2IV. No clear
correlation appeared between the lithium abundance and the rotational
velocity (see also De Medeiros  {\it et al.} 1997). 

In the present work we investigate the physical processes that underline 
the lithium and rotational discontinuities along the subgiant branch. 
For this purpose we first determine the evolutionary status and
individual masses of our sample stars by using the HIPPARCOS parallaxes
and by comparing the observational Hertzsprung-Russell diagram with
evolutionary tracks computed with the Tou\-louse-Geneva code (\S 2). 
This allows us to study very precisely the behaviour of the lithium abundance 
and of the rotational velocity as a function of effective temperature, stellar
mass, metallicity and evolutionary stage. The li\-thium main features are
discussed in \S 3 where we compare the observations with theoretical
predictions related to the dilution mechanism (Iben 1967a,b). Finally, the
connection between the observed rotation behaviour and the magnetic
braking due to the deepening of the convective envelope (Gray 1981, Gray
\& Nagar 1985) is quantified in \S 4. 
This study, based on a close examination of the stellar parameters, 
sheds new light on the question of the link between rotation 
and lithium discontinuities in subgiant stars.

\begin{figure}
\vspace{.2in}
\centerline{\psfig{figure=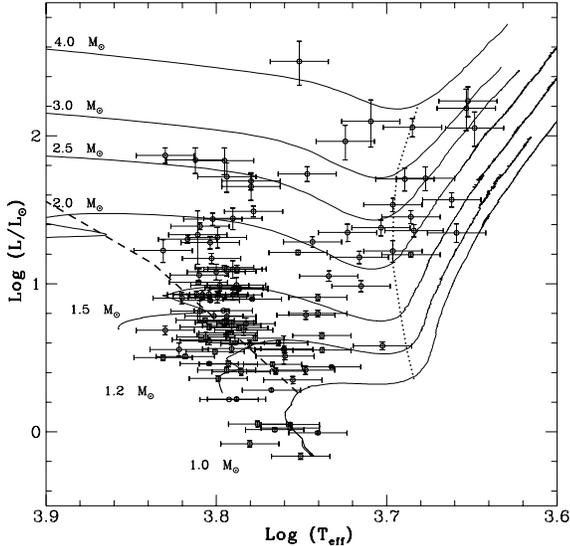,width=3truein,height=3truein}
\hskip 0.1in}
\caption[]{Distribution of the sample stars in the HR diagram.
Luminosities and related errors have been derived from the 
Hipparcos parallaxes.
The typical error on \Teff is  $\pm$ 200~K (Paper~I). 
Evolutionary tracks at [Fe/H]=0 are shown for stellar masses between 
1 and 4~M$_{\odot}$. 
The turnoff and the beginning of the ascent on the red giant branch 
are indicated by the dashed and dotted lines respectively in order to 
discriminate dwarfs, subgiants and giants}
\label{TOa}
\end{figure}

\section{Observational data and evolutionary status}   
\subsection{Working sample}      
 
Our analysis is  based on the observational data from Paper~I.
This sample is composed of 120 Pop~I 
subgiant stars with F, G and K spectral types which belong to 
the ``Catalogue of rotational and radial velocities for evolved stars" 
(De Medeiros \& Mayor 1999), as well as to the Bright Star
Catalogue (Hoffleit and Jaschek 1982). 
We thus use the rotational velocities given in Paper~I, as well as the 
values  derived for \logg, \ALi and \Teff with their respective errors.  

\subsection{Evolutionary status}   
\begin{figure}

\vspace{.2in}
\centerline{\psfig{figure=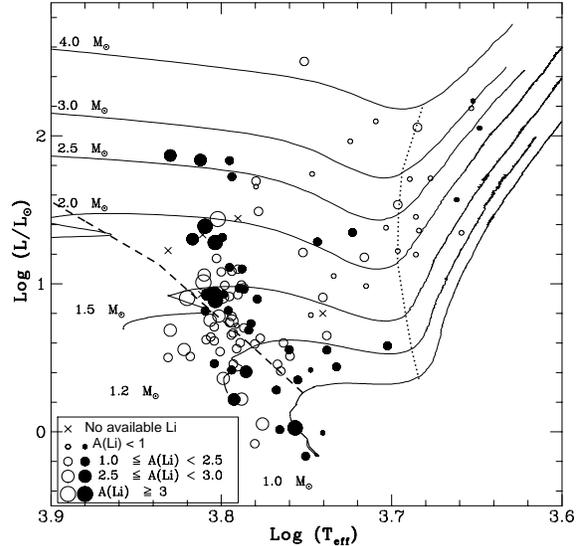,width=3truein,height=3truein}
\hskip 0.1in}

\caption[]{Distribution of the Li abundances in the HRD. Single and binary 
stars are identified with open and filled circles respectively. 
The size of the circles is proportional to the Li abundances quoted in
Paper~I.
Crosses refer to stars without any available Li abundance. 
Evolutionary tracks, as well as the dashed and dotted lines are as in Fig.1}

\label{abunda1}
\end{figure}
In order to interpret  accurately  the observations, we need to know 
the mass and the evolutionary stage of the sample stars. 
We use the HIPPARCOS (ESA 1997) trigonometric parallax measurements to 
locate precisely our objects in the HR diagram.
Among our 120 stars, only one object (HD~144071) has no available Hipparcos 
parallax and is thus rejected from further analysis. 
Intrinsic absolute magnitudes M$_{\rm V}$ 
are derived from the parallaxes and the m$_{\rm V}$ magnitudes 
given by Hipparcos. We determine the  bolometric corrections $BC$ 
by using the  Buser and Kurucz's relation (1992) between  $BC$ and V-I (again 
taken from the Hipparcos Catalogue). Finally, we compute the stellar 
luminosity and the associated error from the sigma error on the 
parallax. The uncertainties in luminosity lower than $\pm0.1$ have an influence
of $\pm0.4$ in the  determination of the masses.
We show the results of these determinations in Fig.~\ref{TOa}  and in
Tab.~\ref{Tabbase}. This table displays the stellar masses and
luminosities derived for all 
 objects of our sample. Moreover complementary data used for the
present analysis ({\it Vsini}, \Teff, \logg, [Fe/H] and  \ALi) can be found 
in Papier I (see Tables 1 to 3). 
The error adopted on T$_{\rm eff}$ ($\pm 200$K) is typical for this class 
of stars, as already discussed in Paper~I. 

We have computed evolutionary tracks with the Toulou\-se-Geneva code
for a range of stellar masses between 1 and 4 M$_{\odot}$ and for 
different metallicities consistent with the range of our sample stars
(see Paper I). However, solar composition being relevant
to most of the objects of our sample (about 65\%), only tracks computed
with [Fe/H]=0 will be displayed in the figures. 
The evolution was followed from the Hayashi fully convective
configuration. We used the radiative opacities by Iglesias \& Rogers
(1996), completed with the atomic and molecular opacities by Alexander
\& Ferguson (1994). The nuclear reactions are from Caughlan \& Fowler
(1988) and the screening factors are included according to the
prescription by Graboske {\it et al.} (1993).
No transport processes except for the classical convective mixing (with
a value of 1.6 for the mixing length parameter) are taken into account.

\subsection {Discrimination between dwarfs, subgiants and giants among
the sample }

In  Fig.~\ref{TOa} we compare the observational HR diagram with 
the evolutionary tracks computed with [Fe/H] = 0.
The dashed line indicates the evolutionary point where the subgiant
branch starts and which corresponds to the hydrogen exhaustion in the 
stellar central regions (i.e., turnoff point).  
About 30 stars are located below the turnoff line and therefore 
appear to be genuine dwarfs, although classified as subgiants in the Bright
Star Catalogue. 
On the other hand, about 15 stars located on the right side of
the dotted line have started the ascent of the RGB and are thus considered 
as giants. 
\begin{table}
\small
\caption{Derived masses and luminosities for our program stars}
\label{Tabbase}
\begin{flushleft}
\begin{tabular}{|r|r|c|}
\noalign{\smallskip}
\hline\\[-2mm]
HD &   $\log(L/L_\odot)$  & M$/$\Msun
\\[2mm]
\hline\\[-3mm]

400     &       0.45$\pm$0.02   &       1.0     \\
645     &       1.36$\pm$0.05   &      1.8     \\
3229    &       0.99$\pm$0.03   &       1.6     \\
3303    &       1.45$\pm$0.04  &       1.5     \\
4813    &       0.21$\pm$0.00   &       1.2    \\
5268    &       1.71$\pm$0.08   &       1.5    \\
6269    &       1.65$\pm$0.08   &       2.4   \\
6301    &       0.64$\pm$0.03   &       1.3     \\
9562    &       0.60$\pm$0.01&  1.5    \\
10142   &       1.69$\pm$0.06   &       2.4    \\
11443   &       1.11$\pm$0.02   &       1.7     \\
12235   &       0.56$\pm$0.03   &       1.2     \\
12583   &       1.70$\pm$0.07&  2.3    \\
13421   &       0.92$\pm$0.03   &       1.4     \\
16417   &       0.45$\pm$0.02   &       1.1        \\
18907   &       0.65$\pm$0.02   &       1.3     \\
19826   &       1.34$\pm$0.05 &        2.2      \\
26913   &       -0.16$\pm$0.02  &       1.1     \\
26923   &       0.05$\pm$0.02   &       1.2    \\
27536   &       1.17$\pm$0.04&  2.2   \\
29613   &       1.49$\pm$0.04   &       2.2    \\
32503   &       2.23$\pm$0.09  &       3.0     \\
33021   &       0.41$\pm$0.02   &       1.1     \\
34411   &       0.28$\pm$0.01   &       1.1     \\
34642   &       1.19$\pm$0.02& 1.8    \\
41700   &       0.22$\pm$0.01   &       1.0    \\
48737   &       1.01$\pm$0.01   &       1.5    \\
60532   &       0.97$\pm$0.02 & 1.3           \\
61421   &       0.81$\pm$0.02 & 1.3     \\
66011   &       0.99$\pm$0.05&  1.5     \\
72954   &       1.05$\pm$0.03   &       1.6     \\
73752   &       0.55$\pm$0.01&  1.2   \\
75487   &       0.90$\pm$0.04&  1.4     \\
78154   &       0.60$\pm$0.02   &1.2    \\
78418   &       0.58$\pm$0.03   &1.2    \\
80956   &       2.09$\pm$0.14 & 3.5     \\
82074   &       0.98$\pm$0.03 & 1.5 \\
82210   &       1.21$\pm$0.01&  1.8    \\
82328   &       0.88$\pm$0.01   &       1.4    \\
82543   &       1.72$\pm$0.09   &       2.4     \\
82734   &       2.05$\pm$0.06   &       3.5        \\
89010   &       0.55$\pm$0.02   &       1.0    \\
89449   &       0.62$\pm$0.01   &       1.3     \\
92588   &       0.59$\pm$0.03   &       1.2     \\
94386   &       1.34$\pm0.06$   &       1.3     \\
99491   &       -0.08$\pm$0.02 &       0.8     \\
100219  &       0.61$\pm$0.03   &       1.3     \\
102713  &       1.27$\pm0.04$   &       1.8     \\
104055  &       2.18$\pm0.14$   &       3.0    \\
104304  &       -0.00$\pm0.00$ &        1.0     \\
104307  &       2.05$\pm$0.11& 2.3     \\
105678  &       1.08$\pm$0.04   &       1.6  \\
107295  &       1.96$\pm$0.11   &       3.5    \\
107700  &       1.86$\pm0.05$   &       2.6             \\
119992  &       0.36$\pm0.02$   &       1.2    \\
120136  &       0.46$\pm$0.01 & 1.2    \\
[2mm]
\hline
\end{tabular}
\\
\end{flushleft}
\end{table}
\begin{table}
(continued)
\begin{flushleft}
\begin{tabular}{|r|r|c|}
\noalign{\smallskip}
\hline\\[-2mm]
HD &   $\log(L/L_\odot)$  &  M$/$\Msun
\\[2mm]
\hline\\[-3mm]
121370  &       0.97$\pm$0.01  &       1.3          \\
123999  &       1.10$\pm$0.03   &       1.6           \\
124570  &       0.74$\pm$0.03   &       1.3     \\
125184  &       0.41$\pm$0.02   &       1.1     \\
125451  &       0.50$\pm$0.02   &       1.3    \\
126400  &       1.22$\pm$0.07&  1.5     \\
126868  &       1.28$\pm0.03$   &       2.2   \\
127243  &       1.74$\pm0.05$   &       2.6    \\
127986  &       1.08$\pm0.05$   &       1.6   \\
130945  &       0.92$\pm0.03$   &       1.5     \\
131040  &       0.68$\pm0.03$   &       1.4     \\
133484  &       0.78$\pm0.03$   &       1.3     \\
136064  &       0.66$\pm0.01$   &       1.3     \\
136202  &       0.70$\pm$0.03&  1.3    \\
137052  &       0.93$\pm$0.02   &       1.5     \\
137510  &       0.68$\pm$0.03   &       1.3    \\
139777  &       0.02$\pm$0.02   &       1.2        \\
142267  &       0.01$\pm$0.02   &       1.1     \\
142980  &       1.56$\pm$0.05&  2.2    \\
144070  &       1.83$\pm$0.08&   2.6     \\
144284  &       0.92$\pm$0.00   &       1.5     \\
144585  &       0.35$\pm$0.02& 1.2   \\
150012  &       1.05$\pm$0.05&  1.6       \\
150680  &       0.89$\pm$0.00   &       1.2    \\
151769  &       1.17$\pm$0.04  &       1.8      \\
154160  &       0.51$\pm$0.03   &       1.1     \\
155078  &       0.92$\pm$0.03   &       1.5    \\
156846  &       0.71$\pm$0.04   &       1.3     \\
157853  &       1.83$\pm$0.09&  2.6     \\
158170  &       1.31$\pm$0.07  &        1.8     \\
161797  &       0.43$\pm$0.00   &       1.1     \\
162003  &       0.75$\pm$0.02&  1.5   \\
162076  &       1.53$\pm$0.05 & 2.6     \\
163989  &       0.91$\pm$0.01   &       1.5   \\
164507  &       0.78$\pm$0.02  &        1.4     \\
172088  &       0.73$\pm$0.04   &       1.3     \\
176095  &       0.91$\pm$0.04   &       1.5       \\
176668  &       1.37$\pm$0.05&  2.3    \\
181096  &       0.73$\pm$0.03   &       1.3     \\
186185  &       0.81$\pm$0.08   &       1.5     \\
190771  &       0.05$\pm$0.00&  1.1     \\
191570  &       0.55$\pm$0.05   &       1.3 \\
196524  &       1.38$\pm$0.02&       1.8        \\
196755  &       0.90$\pm$0.02 & 1.5     \\
196885  &       0.40$\pm$0.02   &       1.2     \\
197373  &       0.50$\pm$0.02   &       1.3 \\
198084  &       0.96$\pm$0.01   &       1.4    \\
199766  &       1.33$\pm$0.17   &       1.9     \\
201507  &       1.22$\pm$0.07  &        1.5     \\
202582  &       0.63$\pm$0.02   &       1.2    \\
206901  &       1.30$\pm$0.03   &       1.6     \\
207978  &       0.54$\pm$0.02   &       1.1     \\
208177  &       1.09$\pm$0.05   &       1.6    \\
210334  &       0.80$\pm$0.02&  1.5    \\
213051  &       1.43$\pm$0.05&  1.7    \\
215648  &       0.65$\pm$0.01   &       1.3     \\
216385  &       0.70$\pm$0.01&  1.3 \\
218640  &       2.50$\pm$0.15   &       4.3          \\
[2mm]
\hline
\end{tabular}
\\
\end{flushleft}
\end{table}

\section{The Lithium main features} 

\begin{figure}

\vspace{.2in}
\centerline{\psfig{figure=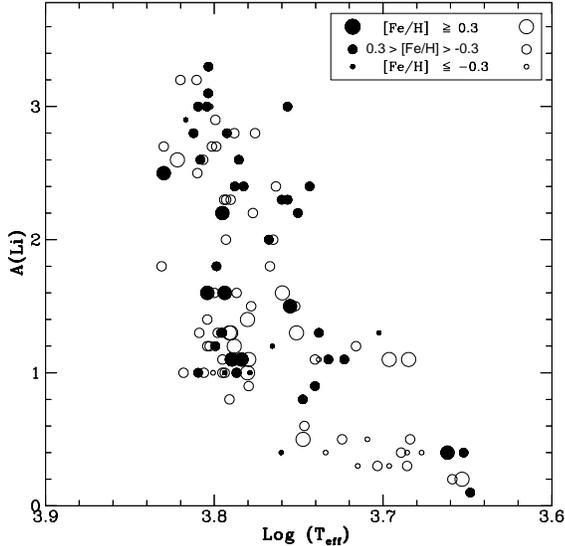,width=3truein,height=3truein}
\hskip 0.1in}

\caption[]{Lithium abundances as a function of  log(T$_{\rm eff}$) 
for all the stars in our sample (Paper~I). Open and
filled symbols represent single and binary stars respectively.
The circle sizes are proportional to the metal content}
\label{metal}
\end{figure}
Fig.~\ref{abunda1} shows the distribution of \ALi on the
HR diagram and Fig.~\ref{metal} 
presents \ALi versus log T$_{\rm eff}$ for three
ranges of metallicity. 
The main features presented in Paper I, i.e., the lithium discontinuity 
around log(T$_{\rm eff}$) equal to 3.75, and the dispersion in lithium 
abundances for subgiants hotter than this value, clearly appear in both 
figures. 
\begin{table}
{\bf Table 1}.(continued)\\
\begin{flushleft}
\begin{tabular}{|r|r|c|}
\noalign{\smallskip}
\hline\\[-2mm]
HD &   $\log(L/L_\odot)$  & M$/$\Msun
\\[2mm]
\hline\\[-3mm]

218804  &       0.4$\pm$0.02  &       1.2            \\
219291  &       1.44$\pm$0.07 &        2.1           \\
219834  &       0.55$\pm$0.09   &       1.2     \\
223346  &       0.78$\pm$0.05   &       1.3     \\ [2mm]
\hline
\end{tabular}
\\
\end{flushleft}
\end{table}

\subsection{Lithium discontinuity} 
\begin{figure}

\vspace{.2in}
\centerline{\psfig{figure=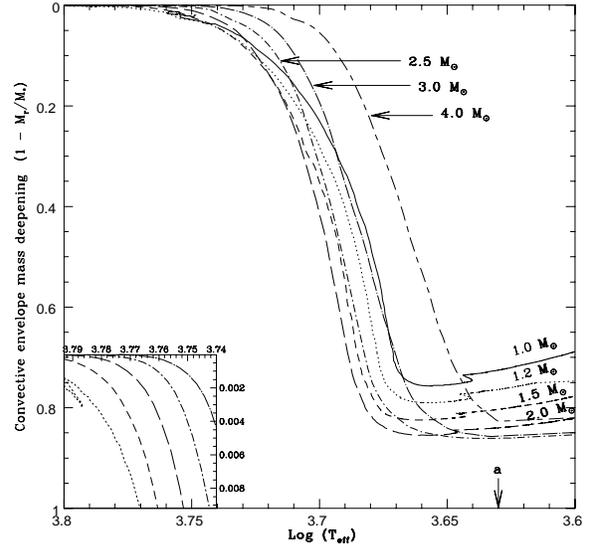,width=3truein,height=3truein}
\hskip 0.1in}

\caption[]{The deepening (in mass) of the convective envelope is shown
as a function of the decreasing effective temperature (first dredge--up) 
for 1.0 (solid), 1.2 (dot), 1.5 (short dash),  2.0 (long dash), 
2.5 (dot - short dash), 3.0 (dot - long dash) and
4.0~M$_{\odot}$ (short dash - long dash ) and [Fe/H] = 0. We present also
a zoom of the region 3.80 $\geq$ \Teff $\geq$ 3.74. 
The point labeled {\it a} indicates the end of the first dredge-up}
\label{bottomb}
\end{figure}
The observational {\bf lithium discontinuity} actually simply reflects the 
well-known dilution that occurs when the convective envelope starts 
to deepen after the turnoff and reaches the inner free-lithium layers 
(Iben 1967a,b). 
In Fig.~\ref{bottomb} we show this behaviour in our models as a function of 
the effective temperature for different masses. A closer look at the models 
shows that the beginning  of the theoretical lithium dilution is a function
of stellar mass, as can be seen in Tab.~\ref{Tb1} where 
{$T^{beg dil}_{\rm eff}$} is the effective temperature at the beginning 
of the lithium dilution for each mass. 
We also give {$T^{f10 dil}_{\rm eff}$}, which is the effective
temperature at which a decrease of the surface lithium abundance by a
factor 10 compared to the value on the main sequence is achieved. 
The dilution is a very fast process, both in terms of age and effective
temperature interval. The theoretical beginning of the lithium dilution 
 is in good agreement 
with the observed abundance drop-off along the subgiant branch, 
as can be seen in Fig.~\ref{4mass}.

For the considered stellar masses, the predicted dilution factor at the end 
of the dredge-up (see point a on Fig.~\ref{bottomb}) ranges between 20 and 60 
(these values obtained from our models do not significantly differ from
the predictions by Iben 1967a,b). 
This corresponds to the upper envelope of the observations on the right side 
of the discontinuity, assuming a cosmic lithium abundance \ALi of 3.1.

\begin{table}
\caption{Relevant characteristics of the  models at [Fe/H]=0 shown 
in  Fig.~\ref{bottomb}. Column 1 gives the stellar masses. 
Columns 2 and 3 give the effective temperature at the point where the
convective envelope just starts deepening and at the onset of the
theoretical lithium dilution. 
 The effective temperature at the point where the lithium abundance
has decreased by a factor 10 compared to its value at the end of the
main sequence is indicated in column 4.  
Column 5 gives the mass of the convective envelope at the 
effective temperature of the observed rotational discontinuity
}

\begin{center}
\tabcolsep0.2cm
\begin{tabular}{c|c|c|c|c}\hline
\multicolumn{1}{c}{mass} &
\multicolumn{1}{c}{Log $T^{dee}_{\rm eff}$ } &
\multicolumn{1}{c}{Log $T^{beg dil}_{\rm eff}$} &
\multicolumn{1}{c}{Log $T^{f 10 dil}_{\rm eff}$} &
\multicolumn{1}{c}{1-$M^{Rot}_{\rm cz}/M_*$}\\ 
\multicolumn{1}{c}{($M_{\odot}$) } &
\multicolumn{1}{c}{} &
\multicolumn{1}{c}{} &
\multicolumn{1}{c}{} &
\multicolumn{1}{c}{} \\ \hline\hline
1.0 &  3.75 & 3.73 & 3.68 &     \\
1.2 &  3.79 & 3.75 & 3.70 & 1.70 $\times 10^{-3}$    \\
1.5 &  3.79 & 3.75 & 3.70 & 4.31 $\times 10^{-4}$     \\
2.0 &  3.77 & 3.74 & 3.72 & 4.22 $\times 10^{-5}$   \\
3.0 &  3.75 & 3.73 & 3.70 & 7.00 $\times 10^{-7}$    \\
4.0 &  3.74 & 3.71 & 3.68 & 8.00 $\times 10^{-7}$   \\
\hline
\end{tabular}
\end{center}
\label{Tb1}
\end{table}

\subsection{Lithium dispersion}
\label{Litdisp}
\begin{figure*}

\vspace{.2in}
\centerline{\psfig{figure=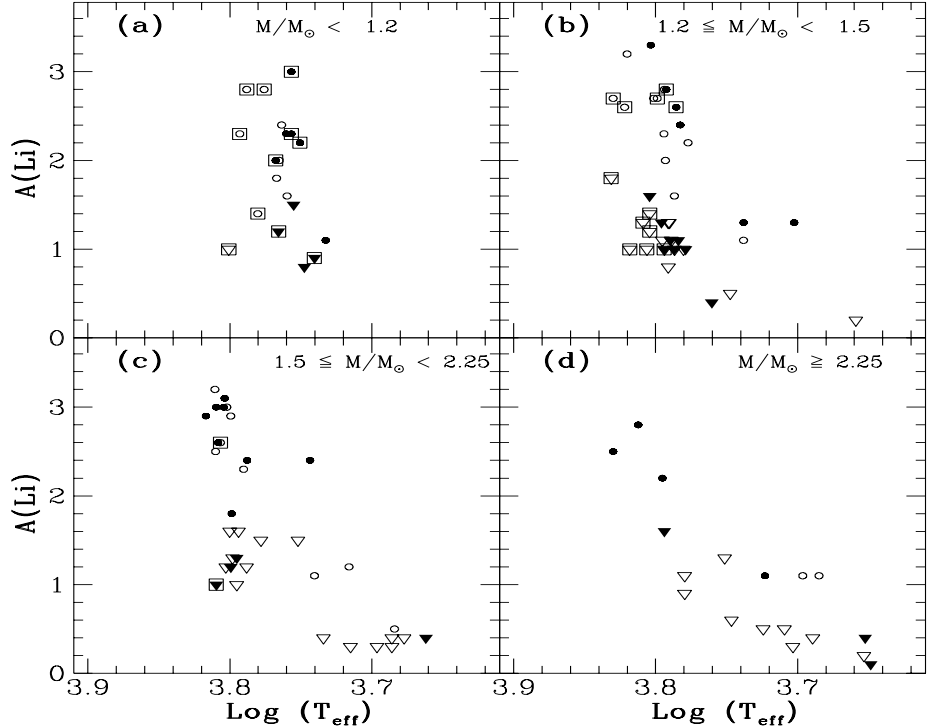,width=5truein,height=4truein}
\hskip 0.1in}

\caption[]{Lithium abundances as a function of  log(T$_{\rm eff}$) for
all our sample stars. Open and filled symbols represent single and binary 
stars respectively.
The circles correspond to lithium detection while inverted triangles
are for upper limits in the  lithium abundance determination. 
Squares point out the main sequence stars}
\label{4mass}
\end{figure*}
In the following, we discuss the {\bf lithium dispersion} on both sides 
of the discontinuity as a function of the stellar mass inferred from the 
evolutionary tracks at the corresponding metallicity.
We select four ranges of mass according to the ones defined in 
Balachandran (1995) for the cluster and field stars with respect to the 
so-called lithium dip region (Boesgaard \& Tripicco 1986). 
Each subsample corresponds to  peculiar observational behaviour of the 
\ALi  on the main sequence (see Fig.~\ref{4mass} where both
main sequence and subgiants stars are shown). In the four mass ranges a
large dispersion of the lithium abundance appears 
and is independent of the single or binary status. 

(i) Stars with masses $<$ 1.2~M$_{\odot}$ show different degrees of
lithium depletion, which occurs already on the pre-main sequence and on 
main sequence, as observed in open clusters and field stars (Soderblom  {\it
et al.} 1993, Jones  {\it et al.} 1999 and references therein). This explains their
low lithium content when they reach the subgiant branch 
(see Fig.~\ref{4mass}a).

(ii) Stars with masses between 1.2 and 1.5~M$_{\odot}$ correspond to the
so-called dip region (Boesgaard \& Tripicco 1986).
 
We separate the stars originating from the hot
side of the dip region in two mass ranges :

(iii) stars with masses between 1.5 and 2.25~M$_{\odot}$, and

(iv) above 2.25~M$_{\odot}$. 

We now discuss in more detail the last three cases. 

\subsubsection{Li-dip stars}
Late F- field and open cluster dwarfs with T$_{\rm eff}$ around 6700 K 
are highly lithium depleted. 
We find this feature among the dwarfs of our sample which have masses 
between $\sim$ 1.2 and 1.5~M$_{\odot}$ (Fig.~\ref{4mass}b), in agreement 
with the observations in the open galactic clusters older than $\sim$ 200 Myr 
(Balachandran 1995).  
This lithium depletion persists in our subgiants of the same mass
interval, in agreement with the observations in slightly evolved stars
of M67 (Pilachowski {\it et al.} 1988, Balachandran 1995, Deliyannis {\it et
al.}
1997). Explanations relying on the nuclear destruction of lithium are thus
favoured by these data (see Talon \& Charbonnel 1998 and references
therein). 

\subsubsection {Li in stars originating from the hot side of the dip}
 
\noindent 
As can be seen in Fig.~\ref{4mass}c,d (see also Fig.~\ref{abunda1}), 
a large lithium dispersion exists among our most massive stars, which 
show lithium depletion by up to two orders of magnitude before the start of 
the dilution at 
 Log T$_{\rm eff} \simeq 3.75$
(point {$T^{beg dil}_{\rm eff}$}). 
Very few observational data are available in the literature 
for stars with masses higher than 1.5~M$_{\odot}$. 
In the Hyades, while on the main sequence these objects show lithium
abundances close to the galactic value, except for a few deficient
Am-stars (Boesgaard 1987, Burkhart \& Coupry 1989).
However, our observational result is in agreement with the findings by 
Balachandran (1990) and Burkhart \& Coupry (1991) of a few slightly evolved 
field stars originating from the hot side of the dip and showing significant 
lithium depletion.
We thus confirm the suggestion by Vauclair (1991, see also Charbonnel \&
Vauclair 1992) that some extra-lithium depletion occurs inside these
stars when they are on the main sequence, even if its signature does not
appear at the stellar surface at the age of the Hyades. 
Among some effects suggested one can quote an unusual mass loss on the main
sequence (Boesgaard {\it et al.} 1977), and rotational induced mixing 
(Charbonnel \& Vauclair 1992, Charbonnel \& Talon 1999).

The behaviour observed in our most massive stars which have not yet reached 
the beginning of dilution explains the very low lithium content of most of 
the massive subgiants which cannot be accounted for by dilution alone. 
The underlying destruction mechanism should also be responsible for the 
lithium behaviour observed in the Hyades giants (Boesgaard {\it et al.} 1977)
which have masses higher than 2.25~M$_{\odot}$ and thus correspond to the
data we show in Fig.~\ref{4mass}d. 
This process should however preserve the boron abundance, which is in 
agreement with the standard dilution predictions in the underabundant Li 
giants of the Hyades, as shown recently by Duncan {\it et al.} (1998).  
Observations of boron in our sample stars have to be done to confirm the
similarity between field and cluster evolved stars.

Finally, we note that Non-LTE effects alone can not explain
the very low Li abundance derived for several massive stars, 
as proposed by Duncan {\it et al.} (1998). Indeed, Carlsson {\it et al.} (1994) have
showed that Non-LTE effects can reach up to ~$\sim +0.3$ dex only for 
such cool stars. 
This correction factor is too low to reconcile the derived Li abundances
with the standard dilution predictions. Therefore, this confirms
that these stars are indeed Li-poor as well as the hotter ones
for which only upper limits have been derived  (Non-LTE
effects are much smaller for such hotter stars).

\section{The {\it Vsini} main features}   

\begin{figure}

\vspace{.2in}
\centerline{\psfig{figure=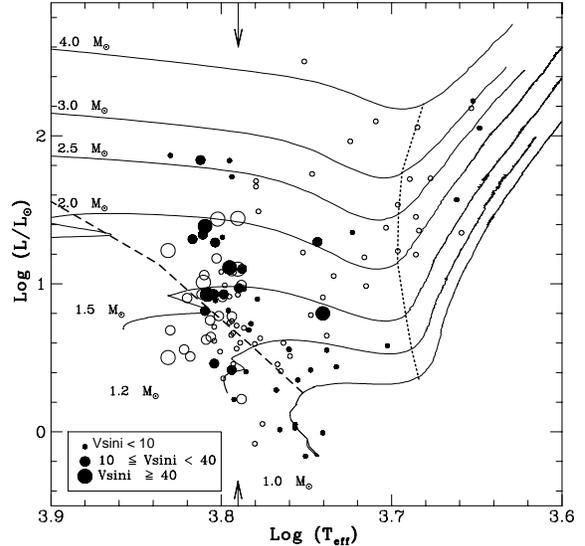,width=3truein,height=3truein}
\hskip 0.1in}
\caption[]{
Rotation for the complete sample. 
The symbol size is proportional to the  rotational velocity measurements
({\it Vsini}, in \kms) obtained with the CORAVEL spectrometer
by De Medeiros \& Mayor (1999).The rotational discontinuity on the
subgiant branch (see Paper~I) is indicated by the two arrows.
Single and binary stars are identified by open and filled circles
respectively}
\label{hr_rot}
\end{figure}
In Fig.6 we present the behaviour of rotation in the  HR
diagram for our sample stars, including the dwarfs and giants. 
We define several {\it Vsini} intervals as in L\`ebre {\it et al.} (1999): 
Slow rotators correspond to {\it Vsini} $<$ 10~\kms, moderate rotators 
to 10~\kms $\leq$ {\it Vsini}  $<$ 40~\kms, and high rotators to 
{\it Vsini} $\geq$ 40\kms. 
Fig.6 shows clearly the now well established rotational 
discontinuity along the subgiant branch near log(T$_{\rm eff}$)=3.79, 
here indicated by two arrows (Gray \& Nagar 1985; De Medeiros \& Mayor 
1989, 1990).
An interesting feature to discuss is the influence of stellar mass on the 
rotational discontinuity.
First of all, one observes that all the subgiant stars with mass lower
than about 1.2~M$_{\odot}$ present {\it Vsini} values lower
than 10.0~\kms. Observations in young galactic clusters (see Gaig\'e 1993 
and references therein) show that these low-mass stars actually acquire a 
slow rotation early on the main sequence, as explained by the magnetic
braking scenario for main sequence stars (Kraft 1967,  Schrijver and Pols 1993). 

Before the rotational discontinuity, the subgiants with mass larger than 
1.2~M$_{\odot}$ present a broad range of {\it Vsini} values. 
Because their very thin surface convective envelopes are not an
efficient site for magnetic field generation via a dynamo process, 
these stars are not expected to experience significant angular momentum loss
during their main sequence evolution. This result is again in agreement
with the data in open clusters.

As can be seen in Tab.~\ref{Tb1} (see also Fig.~\ref{bottomb}), 
the effective temperature at which the convective envelope starts to
deepen (T$^{dee}_{\rm eff}$) depends slightly on the stellar mass. 
We also give the depth of the convective envelope at the effective
temperature of the rotational discontinuity, M$^{Rot}_{\rm cz}$. \
 For the masses lower than 2.0 M$_{\odot}$, the observed rotation
discontinuity occurs just when the convective envelope starts deepening.
We thus see that if the magnetic braking plays a
relevant role in the rotational discontinuity, it requires only 
a very small change in the mass of the convective envelope. 
Above 2.0~M$_{\odot}$, our sample is very sparse (due in particular to
the very rapid evolution of such stars in the Hertzsprung gap) and we 
have no data for
single stars on the left of the rotation discontinuity exhibited by 
lower stellar masses. We cannot thus discuss further the impact of the
deepening of the convective envelope on the braking in these more
massive stars. 

\section{Conclusions}

In this work we have  analyzed the Li and rotation observations for
120  subgiant stars from  L\`ebre {\it et al.} (1999). 
We used the HIPPARCOS  trigonometric parallax measurements to
locate precisely our objects in the HR diagram and to 
determine the individual mass and evolutionary status for all stars in 
the sample.

We have compared  observed Li abundances with predictions of Li dilution 
caused by the deepening of the convective envelope on the subgiant branch.
Our models show that the beginning  of the theoretical lithium dilution 
is a function of stellar mass and coincides with the observational features. 
Stars with masses $<$ 1.2~M$_{\odot}$  show a large range in abundance
before the turnoff, indicating lithium depletion in the previous phases. 
Stars with masses between 1.2 and 1.5~M$_{\odot}$ (i.e., in the dip region) 
show  \ALi values in agreement with what is found in the open clusters. 
We note that many stars with masses higher than 1.5~M$_{\odot}$
show lithium depletion up to two orders of magnitude before the start of the
dilution at Log T$_{\rm eff} \simeq 3.75$. 
The process that depletes Li in these objects while on the main sequence 
should however preserve the boron abundance, which is in
agreement with the standard dilution predictions in the underabundant Li
giants of the Hyades, as shown recently by Duncan {\it et al.} (1998).
Because these specie burn at different depths than lithium, future 
observations of boron for our sample  will 
provide powerful additional constraints  and will confirm the
similarity between field and cluster evolved stars. 

Our analysis 
 confirms
that low mass stars leave the main sequence with a
low rotational rate, while more massive stars are slowed only when
reaching the subgiant branch. A very slight increase of the depth of the
convective envelope seems to be sufficient for the magnetic braking
to take place  at this phase.
 Even  if the  decrease in the behavior of Lithium is  nearly
the rotational discontinuity, our  interpretation  of the
observations  shows that  lithium and rotation discontinuities are independent.

\begin{acknowledgements}
J.D.N.Jr. acknowledges partial financial support from the CNPq Brazilian
Agency. PdL acknowledges  partial  support  grants  from the
{\it Soci\'et\'e de Secours des Amis des Sciences} and
the {\it  Fondation des Treilles}.

\end{acknowledgements}                 

{}
\end{document}